\newcommand*{\memfontfamily}{pazo}
\newcommand*{\memfontpack}{mathpazo}

\documentclass[10pt,
onecolumn,oneside,a5paper,article
,frenchb,italian,german,swedish
,british%
]{memoir}

\usepackage[mathpazo
,eugreek
,optima
,itsans
,lmodern
]{memmana2}

\hypersetup{%
breaklinks=true,%
pdfauthor={PierGianLuca  Porta Mana},
pdftitle={In favour of the time variable in classical thermoDYNAMICS}
}

\usepackage{mathdots}

\renewcommand{\multicitedelim}{\addsemicolon\space}
\bibliography{manabibliography}
\renewcommand{\bibfont}{\small}

\DeclareBibliographyCategory{1}
\newcommand{\citep}{\parencites}
\newcommand{\citey}{\parencites*}
\renewcommand{\cite}{\citep}

\setlxvchars
\setxlvchars
\settypeblocksize{*}{\lxvchars}{1.618}
\setulmargins{*}{*}{*}
\setlrmargins{*}{*}{*}
\setheaderspaces{*}{*}{*}
\checkandfixthelayout

\providecommand{\firstname}[1]{#1}
\providecommand{\surname}[1]{#1}
\providecommand{\affiliation}[1]{{\footnotesize#1}}
\providecommand{\epost}[1]{{\footnotesize #1}}
\providecommand{\email}[1]{{\textsf{\href{mailto:#1}{#1}}}}
\providecommand{\pacs}[1]{{\footnotesize\textsc{PACS} numbers: #1}}
\providecommand{\msc}[1]{{\footnotesize\textsc{MSC} numbers: #1}}

\ifdraftdoc
\usepackage
{datetime}
\newdateformat{mydate}{\THEDAY\ \monthname[\THEMONTH] \THEYEAR}
\fi

\newcommand{\asudedication}[1]{%
{\par\centering\textit{#1}\par}}

\newenvironment{acknowledgements}{\section*{Acknowledgements}\addcontentsline{toc}{section}{Acknowledgements}}{\par}

\chapterstyle{reparticle}
\renewcommand*{\thesection}{\arabic{section}}
\copypagestyle{manaart}{plain}
\makeheadrule{manaart}{\headwidth}{0.5\normalrulethickness}
\makeoddhead{manaart}{%
\footnotesize\textit{Porta Mana}}{}{
\footnotesize\textit{Time variable in thermo\emph{dynamics}}} 
\makeoddfoot{manaart}{}{\thepage}{}
\makeatletter
\newcommand\addprintnote{%
\begin{picture}(0,0)%
\put(0,-14){
\makebox(0,0){
{\tiny%
This document is optimized for on-screen reading and 2-pages-on-1-sheet
printing on A4 or Letter paper}}
}%
\end{picture}%
}
\makeoddfoot{plain}{}{\makebox[0pt]{\thepage}\addprintnote}{
}
\makeoddhead{plain}{}{}{\footnotesize
Report no.\ pi-other-204
}

\title{In favour of the time variable in\\ classical thermo\emph{dynamics}
}
\author{\firstname{PierGianLuca} \surname{Porta\,Mana}
\\
\affiliation{Perimeter Institute for Theoretical Physics, 
Canada}
\\[-1ex]
\epost{\textless\email{lmana@perimeterinstitute.ca}\textgreater
}%
}

\predate{\begin{center}\footnotesize}
\date{19 January 2011
{
\\ (first drafted 29 October 2010)}
}
\postdate{\end{center}}

\usepackage{breakurl} 

\newcommand{\zrc}{\rho_0}
\newcommand{\zta}{\tau}
\newcommand{\zla}{\lambda}
\newcommand{\zte}{\theta}

\newcommand{\zpi}{\pi}
\newcommand{\ztt}{\bm{\mathsf{T}}}
\newcommand{\zT}{T}
\newcommand{\zD}{\bm{\mathsf{D}}}
\newcommand{\zq}{\bm{q}}
\newcommand{\zv}{\bm{v}}
\newcommand{\zF}{\bm{\mathsf{F}}}
\newcommand{\zr}{\rho}
\newcommand{\zg}{f}
\newcommand{\zM}{M}
\newcommand{\zti}{t_0}
\newcommand{\ztp}{\tau_\text{o}}
\newcommand{\zQ}{\varPhi}
\newcommand{\zW}{P}
\newcommand{\zqq}{\phi}

\newcommand{\zLa}{\lambda}
\newcommand{\ztaa}{\zta}
\newcommand{\zLh}{L_\text{r}}

\newcommand{\ztV}{V_\text{v}}

\newcommand{\zRi}{\Bar{R}}

\newcommand{\zhh}{\Bar{h}}
\newcommand{\zacc}{a}
\DeclareMathOperator{\dive}{div}

\begin{document}
\selectlanguage{british}


\maketitle
\abslabeldelim{:\quad}
\setlength{\abstitleskip}{-\absparindent}
\abstractrunin
\begin{abstract}
  A case for the teaching of classical thermodynamics with an explicit time
  variable, with phenomena involving changes in time, is made by presenting
  and solving a exercise in textbook style, and pointing out that a
  solution accords with experiment. The exercise requires an explicit
  treatment of the time variable. Further arguments are given for the
  advantages of an explicit time variable in classical thermodynamics, and
  against some standard terminology in this theory.
  \\[2\jot]
  \pacs{05.70.-a,64.10.+h,01.40.gb}\\ \msc{80-01,97M50,34-04}
\end{abstract}
\asudedication{to Laura}

\setlength{\epigraphrule}{0pt}
\epigraph{`i mean 
if you dont change in 2 years or even in a day, then whats the point?'}%
{\textit{L. Pasichnyk}}

\newrefsegment
\selectlanguage{british}

\newcommand{\figsol}{\begin{figure}[!p]
  \centering
\includegraphics[width=0.59\columnwidth]{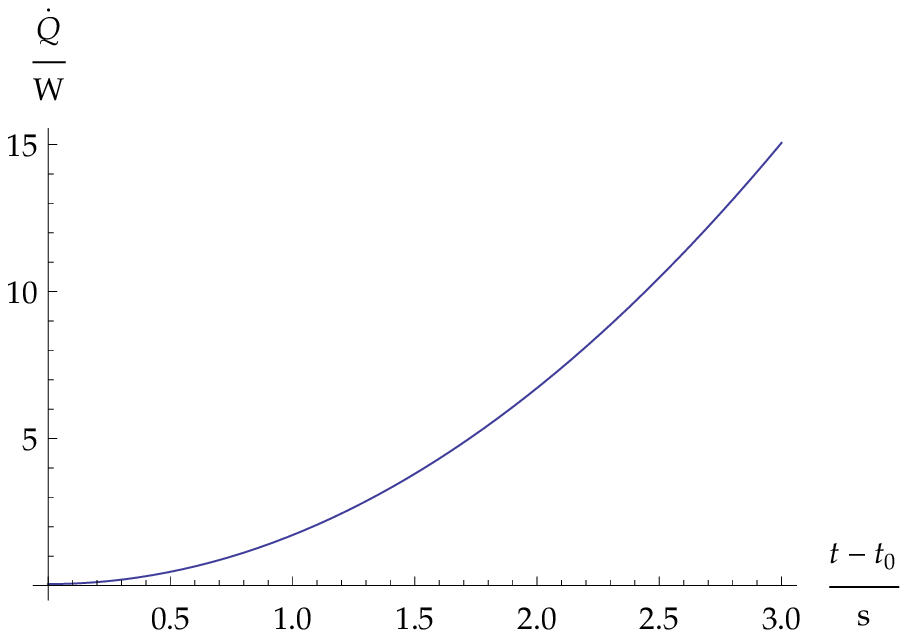}
\includegraphics[width=0.59\columnwidth]{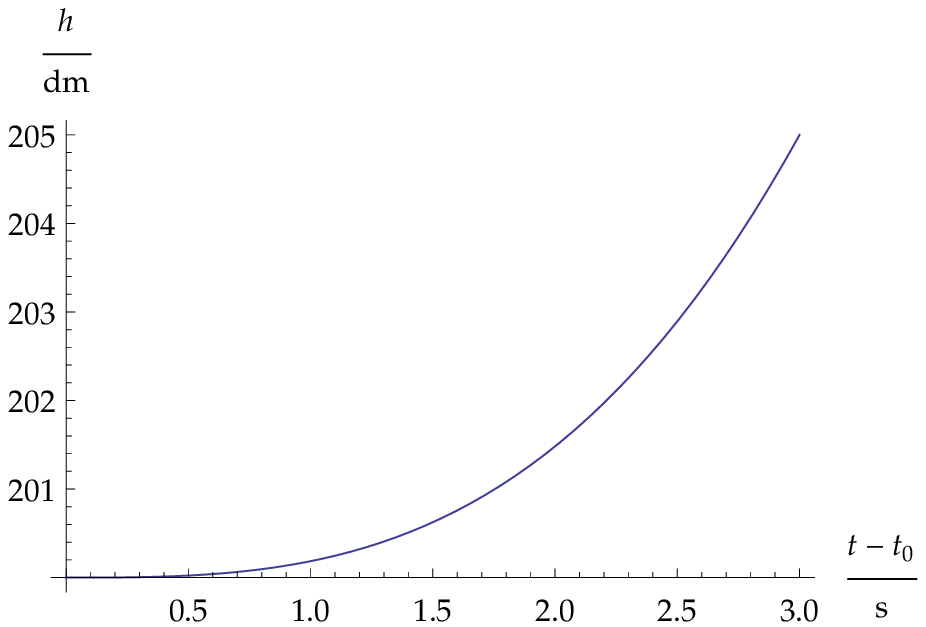}
\includegraphics[width=0.59\columnwidth]{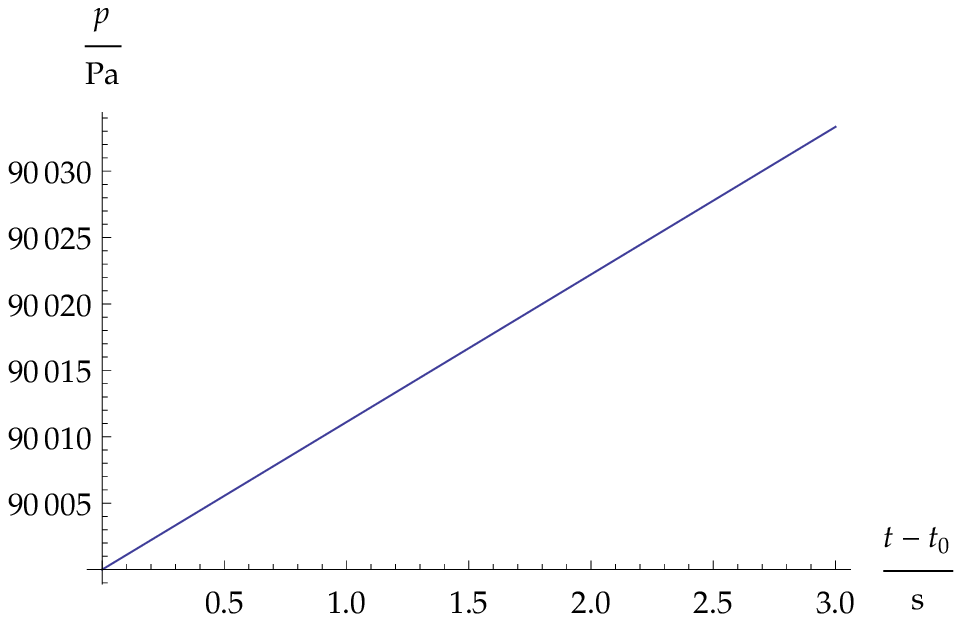}
\includegraphics[width=0.59\columnwidth]{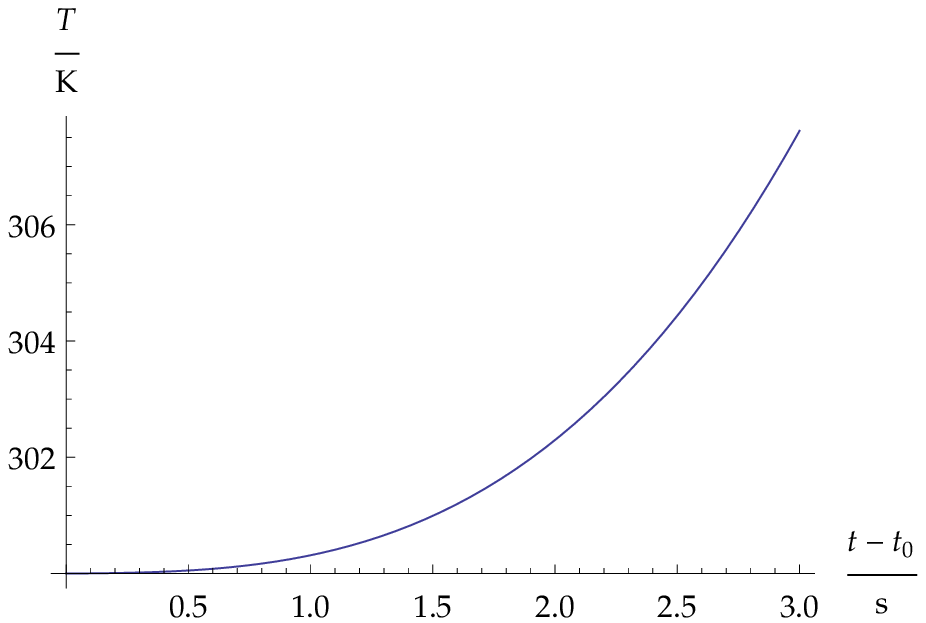}
\caption{Solution for problem~\ref{item:findQ}: $h(t) = L + \zLa
  \frac{(t-\zti)^3}{27\ztaa^3}$ and the numerical
  values~\eqref{eq:num_values}}
\label{fig:h1}
\end{figure}}

\newcommand{\fige}{\begin{figure}[!p]
  \centering
\includegraphics[width=0.59\columnwidth]{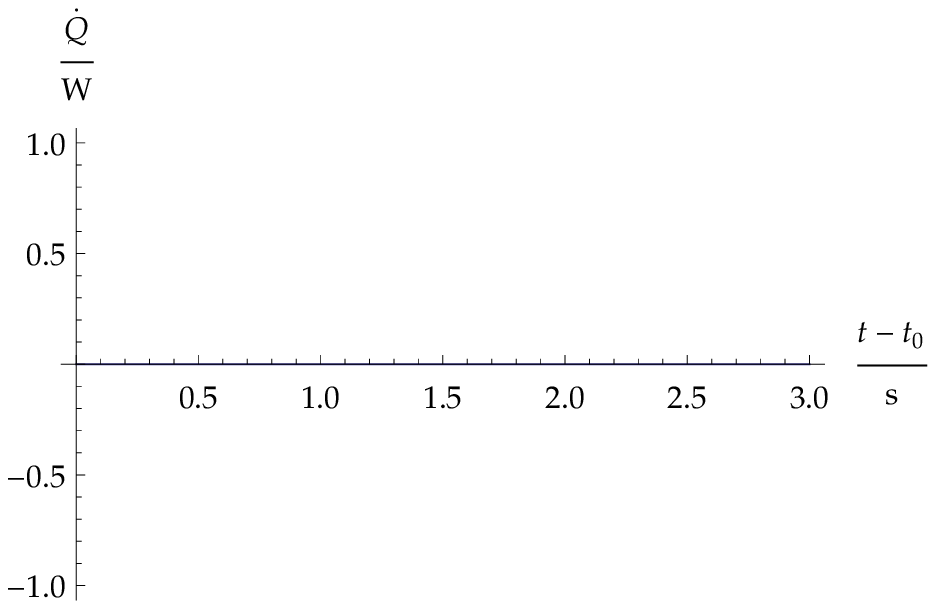}
\includegraphics[width=0.59\columnwidth]{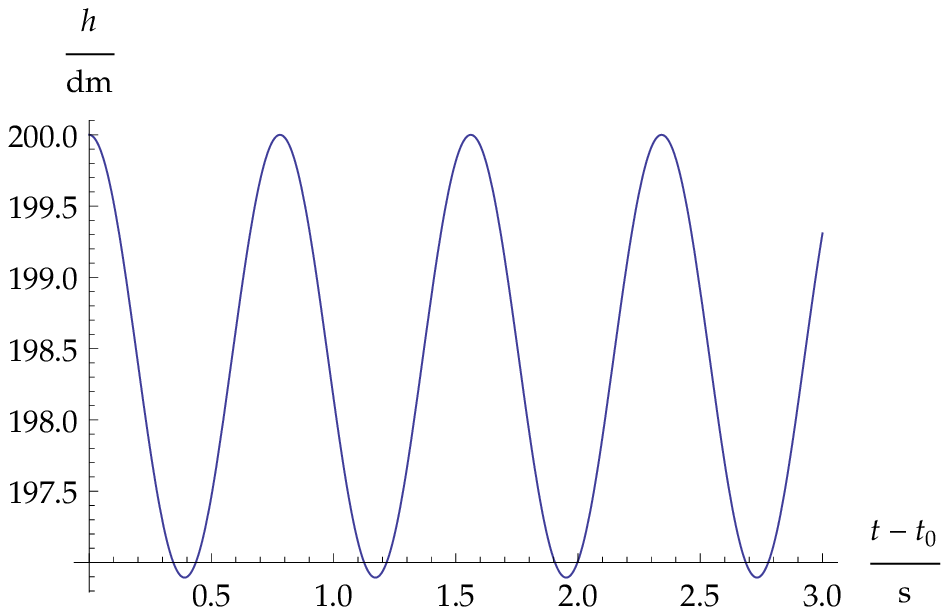}
\includegraphics[width=0.59\columnwidth]{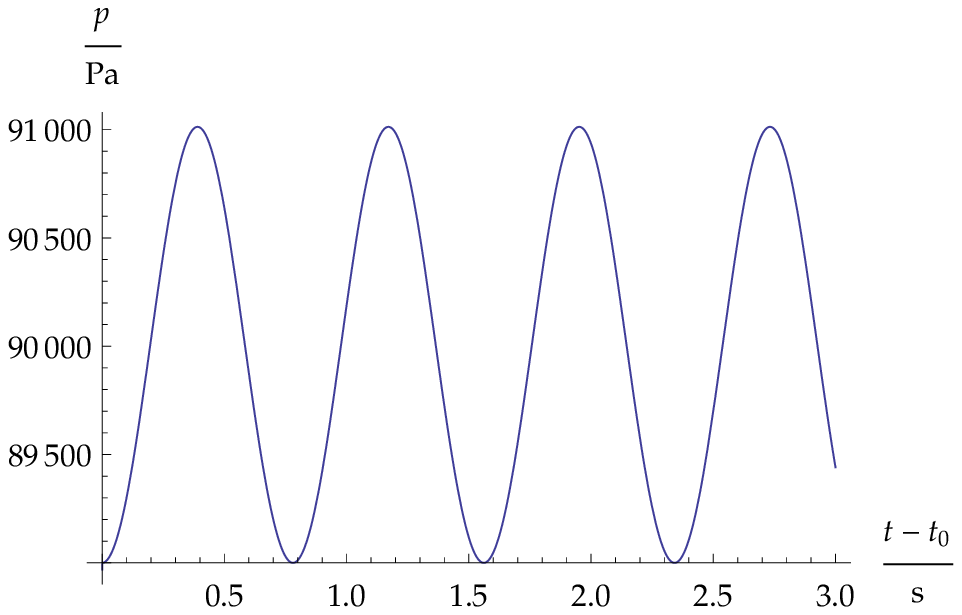}
\includegraphics[width=0.59\columnwidth]{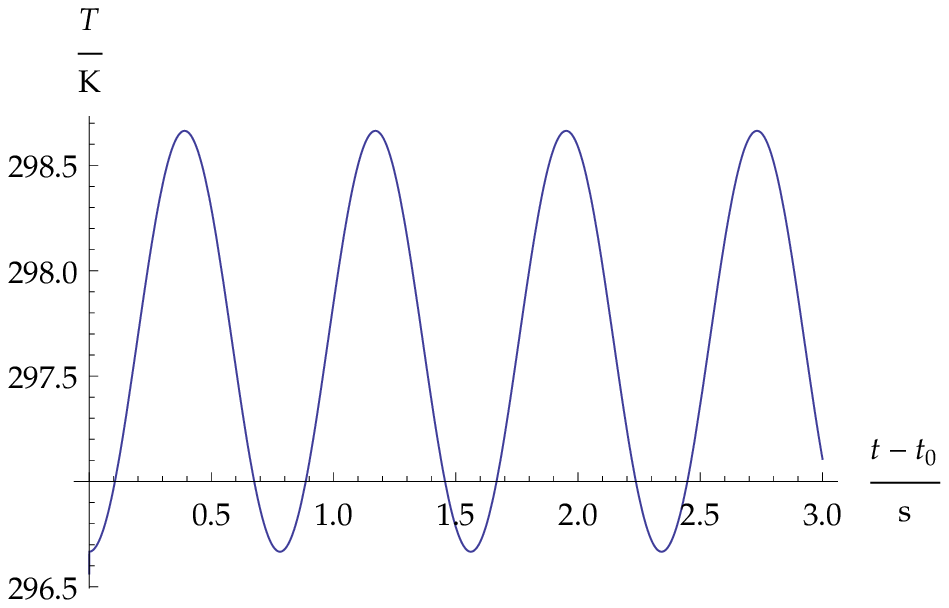}
\caption{Solution for problem~\ref{item:findh}: $\zQ(t)=0\un{W}$, an
  initial negative acceleration of $\ddot{h}(\zti)=\zacc$, and the
  numerical values~\eqref{eq:num_values}}
\label{fig:h02}
\end{figure}}

\newcommand{\figdiff}{\begin{figure}[!p]
  \centering
\includegraphics[width=0.77\columnwidth]{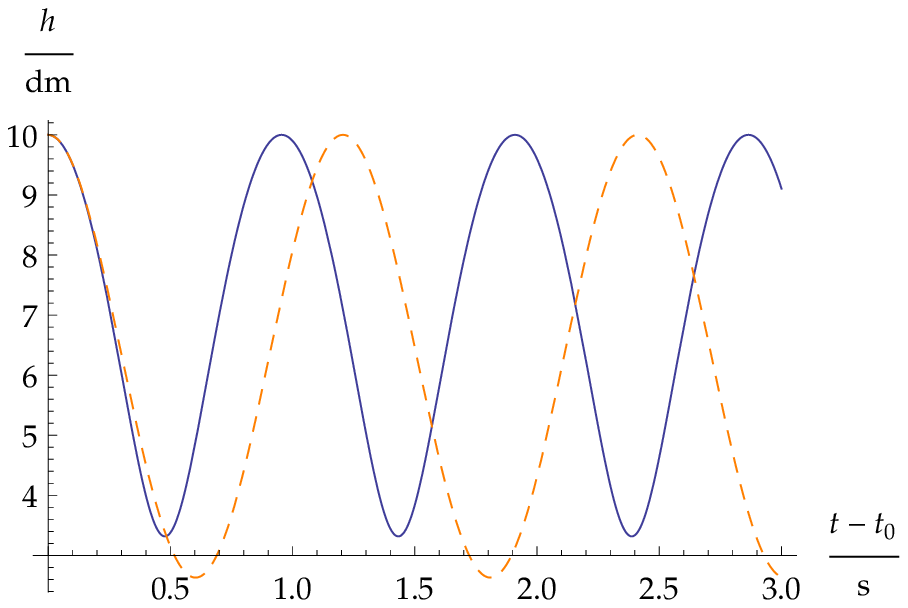}
\includegraphics[width=0.77\columnwidth]{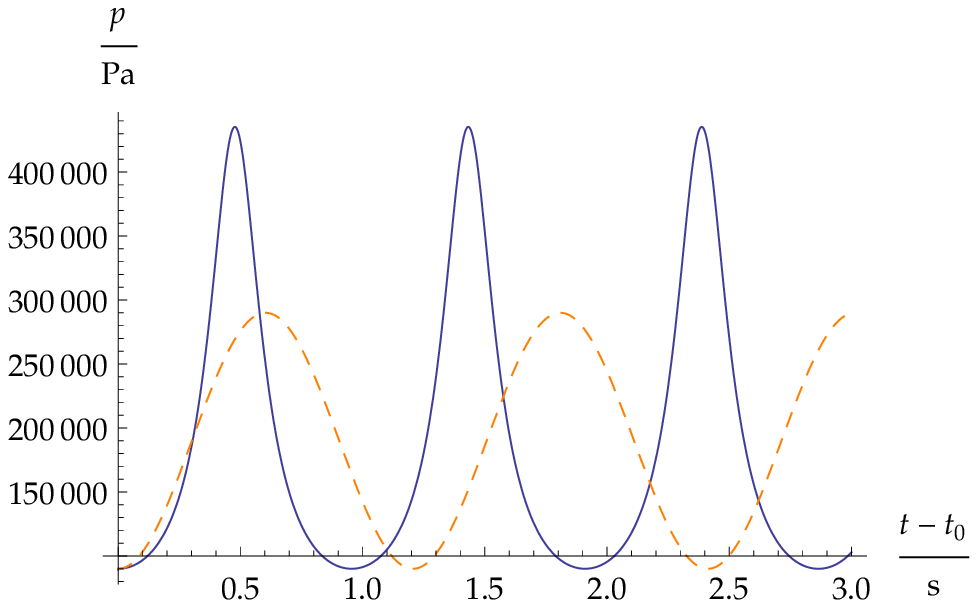}
\includegraphics[width=0.77\columnwidth]{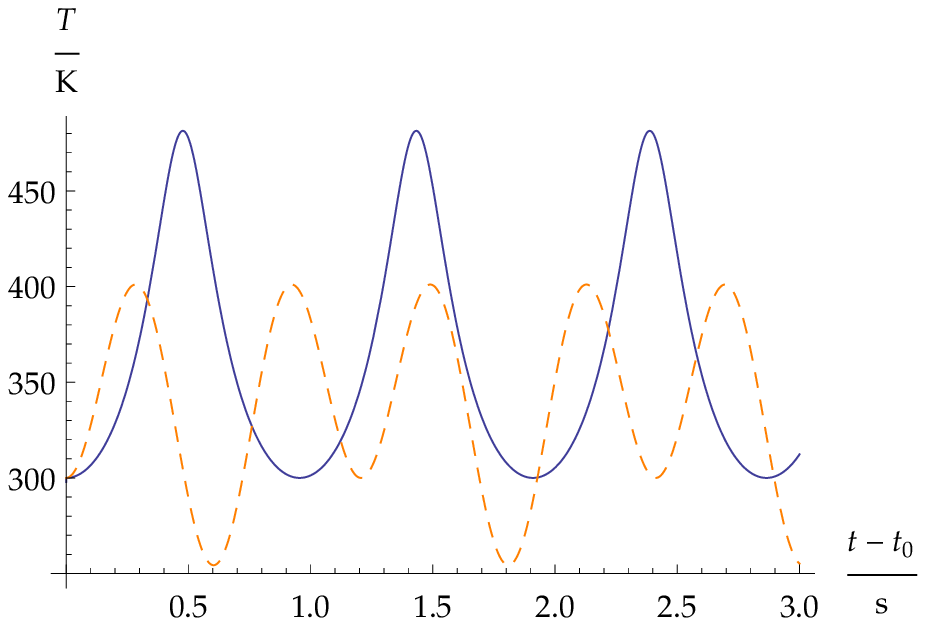}
\caption{Comparison of solutions for problem~\ref{item:findh} with the
  numerical values~\eqref{eq:num_values_new}, using the exact
  equation~\eqref{eq:sol_Q_h} (continuous blue line) and the approximate
  equation~\eqref{eq:approx_harm} (dashed orange line)}
\label{fig:hd}
\end{figure}}

\newcommand{\figset}{\begin{figure}[!btp]
  \centering
\includegraphics[width=1\columnwidth]{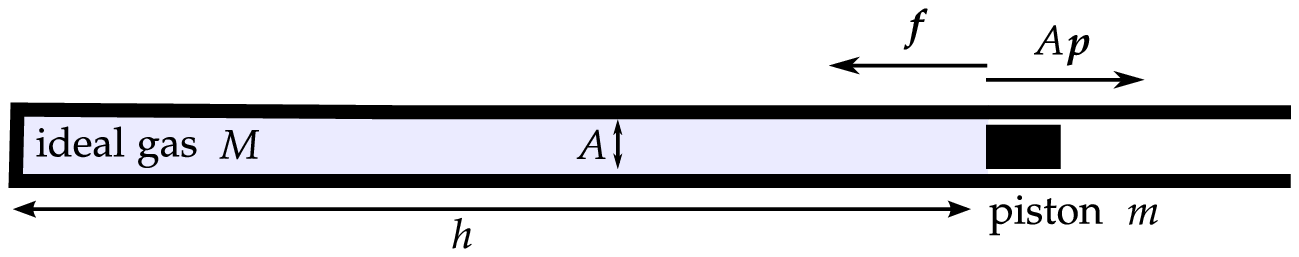}
\caption{Idealization of the experimental set-up of fig.~\ref{fig:schufle}}
\label{fig:cylinder0}
\end{figure}}

\newcommand{\figass}{\begin{figure}[!tbp]
  \centering
\includegraphics[width=0.8\columnwidth]{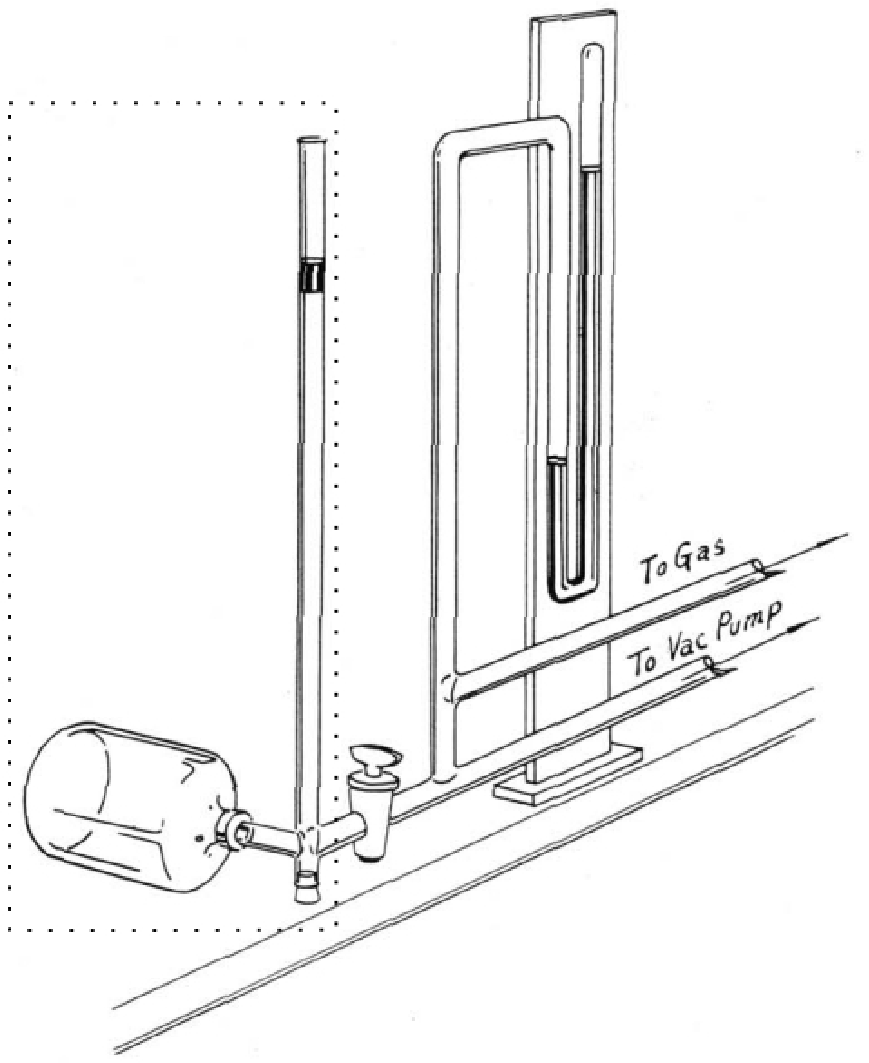}
\caption{Assmann-R\"uchardt-Schufle apparatus. The part of the apparatus
  inside the dotted rectangle is idealized in fig.~\ref{fig:cylinder0};
  notice the piston in the left tube. With this apparatus the solution of
  problem~\eqref{eq:h_question}, shown in fig.~\ref{fig:h02}, can be
  experimentally observed. (Reprinted and adapted from Schufle
  \citey{schufle1957} with the permission of the American Chemical
  Society)}
\label{fig:schufle}
\end{figure}}

\section{Introduction}
\label{sec:intro}

The way I learnt classical thermodynamics in my undergraduate studies was
very different from the way I learnt mechanics and electromagnetism. The
most important part of these two disciplines was about changes in time and
the laws governing these changes; in fact, their fundamental laws are
differential equations involving time derivatives. In thermodynamics,
instead, there was no change; all exercises were mainly about static
equilibrium situations and their comparison: the fundamental laws of
thermodynamics did not involve time and were stated by means of
differentials, not derivatives.

This difference in philosophy and mathematics was frustrating. First of
all, I had a feeling of a lack of unity in Physics. For mechanics and
electromagnetism could easily be made to interact with each other, proving
to be only two simplified aspects of that variety of physical phenomena
that every day fills our senses with marvel. Yet, our senses also tell us
that this variety includes thermal phenomena; but this was not reflected in
classical thermodynamics: its interaction with the other two disciplines
was scarce and limited to trivial problems.

Time and change were not completely absent, of course. `Quasistatic'
processes were named (with a lot of verbal exorcisms owing to their
`non-reality'), but never developed mathematically with the introduction of
a time variable. Time only appeared explicitly in very elementary problems
of heat conduction that involved heat and temperature only, or of
diffusion.

I believe my experience is common to many students, especially in
departments of physics, as can be evinced by a look at thermodynamic
textbooks. For example, in \sect~2.1, p.~30, of the widely used book by
Zemansky \citey{zemanskyetal1937_r1997} we read:
\begin{quote}
Thermodynamics does not attempt to deal with any problem involving the 
rate at which a process takes place. The investigation of problems involving 
the time dependence of changes of state is carried out in other branches of 
science, as in the kinetic theory of gases, hydrodynamics, and chemical 
kinetics.
\end{quote}
Or in Wilson's \citey[\sect~1\textperiodcentered1, p.~2]{wilson1957}:
\begin{quote}
  Another way in which thermodynamics differs from other subjects is 
that it only deals with equilibrium states and with transitions from one 
equilibrium state to another. 
\end{quote}
Callen \citey[\sect~1-1, p.~6]{callen1960_r1985} says, more concisely but in
italics:
\begin{quote}
  \emph{
thermodynamics describes only static states of
    macroscopic systems.}
\end{quote}

There are exceptions. There are textbooks in which the fundamental
principles of thermodynamics are explained and expressed from the beginning
in terms of change and time; for example Sandler's \citey[see \chaps~3,
4]{sandler1977_r2006}, Moran \amp\ Shapiro's \citey[see \chaps~2,
6]{moranetal1988_r2006}, and Bohren \amp\ Albrecht's refreshing textbook
\citey{bohrenetal1998} (see also Winterbone's
\citey[passim]{winterbone1997}). One can also find an occasional article,
like Murphy's \citep{murphy1979}, where thermodynamic variables are treated
with an explicit time dependence without reference to kinetic theory or
`irreversible thermodynamics' (I do not count exercises about heat
conduction as they typically involve heat and temperature only). So there
are courses in which thermodynamics is taught that way; such courses seem
to be more numerous in departments of engineering and chemistry than of
physics.

Why is thermodynamics taught without the time variable? and in which kind
of courses and departments is it taught differently? It is not my purpose
here to answer these questions; for the first, you may take a look at the
researches by Truesdell \etal\
\citep{truesdell1969_r1984,truesdelletal1977,truesdell1980,truesdell1986b}.
Rather, I should like to show, to those students and teachers who have had
experiences similar to mine, that thermodynamics can be approached in a way
more similar to mechanics and electromagnetism, with change and time. Cases
for the explicit appearance of time in thermodynamics were already made by
Eckart \citey{eckart1940a}; by Bridgman \citey[see \eg\
pp.~24--26]{bridgman1941_r1943}; extensively and intensively by Menger
\citey{menger1950}; very forcefully, with telling historical, physical, and
mathematical arguments, by Truesdell and his school of `rational
thermodynamics' (see \eg\
\citep{truesdelletal1960,truesdelletal1965_r2004,truesdell1966b,truesdell1968c,truesdell1969_r1984,truesdelletal1977,truesdell1980,owen1984,samohyl1987,day1988,ericksen1991_r1998,silhavy1997});
and by Bohren \amp\ Albrecht \citey[\sects~1.8, 3.1]{bohrenetal1998}. Here
I will show this time-dependent thermodynamics in action by proposing and
solving a thermodynamic exercise in a typical textbook style. Note that the
solution of the exercise can be and is experimentally verified in a simple
undergraduate laboratory experiment, as will be discussed later in
\sect~\ref{sec:solution}; so there is no arguing that the exercise has no
physical meaning or application. The exercise is suited to students with a
knowledge of differential equations, but could be re-expressed in a simpler
way that avoids direct mention of differential equations (as is done with
the harmonic oscillator, \eg, in basic courses).

Afterwards I will argue that the presentation of thermodynamics with an
explicit time dependence and time derivatives is more advantageous,
more general, and physically, mathematically, philosophically more appealing
that the one based on the absence of time and the mathematics of
differentials.

\subsection{Classical thermodynamics with time}
\label{sec:TDwithtime}

Before presenting the exercise, let us see how the time variable is
introduced in classical thermodynamics. The change is very simple:
\begin{asparaitem}
\item The variables describing the system depend on time: the independent
  ones directly, and the dependent ones indirectly through their dependence
  on the independent variables. For example, if volume $V$ and temperature
  $\zT$ are the independent variables, and pressure $p$, internal energy
  $U$, entropy $S$ the dependent ones, we shall have $V(t)$, $\zT(t)$,
  $p[V(t), \zT(t)]$, $U[V(t), \zT(t)]$, \etc
\item The differentials usually appearing in the first and second laws are
  recognized to be differentials with respect to time, and therefore
  replaced by time derivatives. So the first and second law assume the forms
  \begin{equation}
    \label{eq:laws_time}
    \dot{U}=\zQ -\zW,\qquad
\dot{S} \ge \frac{\zQ}{\zT}.
  \end{equation}
  In these equations, $\dot{U}\defd\di U/\di t$ and $\dot{S}\defd \di S/\di
  t$ are the rates of change of internal energy and entropy, $\zQ$ is the
  \emph{heating}, or rate at which the system is heated, and $\zW$ is the
  \emph{working}, or rate at which the system does work. The latter is
  usually specified in terms of the independent and dependent variables of
  the system and their rates of change, \eg\ $\zW=p\dot{V}$. (The term
  `power' is usually reserved for the sum of the working and the rate of
  change of kinetic energy, the latter being often negligible in
  thermodynamics.)
\end{asparaitem}

This formulation was discussed by Menger \citey{menger1950} and is
presented in Truesdell
\citey[Lecture~26]{truesdell1966b}{truesdell1968c}[Lecture~1 and
Appendix1A]{truesdell1969_r1984}, Truesdell \amp\ Bharatha
\citey{truesdelletal1977}, Owen \citey{owen1984}, Samoh\'yl
\citey[\chap~II]{samohyl1987}, and the already mentioned Sandler
\citey{sandler1977_r2006}. See also M\"uller
\citey[\sect~5.1]{mueller1985}.

\bigskip

Now to our exercise:

\section{A textbook exercise in thermo-\emph{dynamics}}
\label{sec:genintro}
A vessel communicating with a tube as in the set-up of
fig.~\ref{fig:schufle} is filled with a mass $\zM$ of an ideal gas. This
set-up can be idealized as a tube of constant cross-section area $A$ as in
fig.~\ref{fig:cylinder0}: \figass \figset
the gas is enclosed between the closed base of the tube and a piston of
mass $m$ that fits the tube and can move along it without friction. The
variable distance of the piston from the base is denoted by $h(t)$. A
constant force $f$ acts on the piston from the outside (the force can \eg\
consist in atmospheric pressure, and also in the weight of the piston
itself if the tube is in a vertical position). The gas is homogeneous at
all times, and is characterized by an individual massic gas constant $R$
(\ie\ $R=\zRi/M_\text{m}$ with $\zRi$ the ideal-gas constant and
$M_\text{m}$ the molar mass of the gas) and a linear relation between
internal energy and temperature with massic heat capacity $c$. Consider the
following two distinct problems:
\begin{asparaenum}[A.]
\item\label{item:findQ}Until time $\zti$ the piston is at a distance $L$ from the base of
  the tube and the gas is in equilibrium. Starting at time $\zti$ heating
  is suddenly provided thereafter at a rate $\zQ(t)$. How must
  this heating rate depend on time if the distance of the piston from its
  initial position has to change like
\begin{equation}
  \label{eq:h_question}
  h(t) = L + \zLa \frac{(t-\zti)^3}{\zta^3}
\end{equation}
with $\zLa$ and $\zta$ two constant positive parameters? How do pressure
and temperature change?

\item\label{item:findh}At time $\zti$ the piston is at a distance $L$ from
  the base of the tube, with nought velocity but an initial negative (\ie,
  towards the base) acceleration $\zacc$. The gas is adiabatically
  insulated at all times, \ie\ the heating vanishes, $\zQ(t)=0$. What is
  the motion of the piston after time $\zti$? If necessary, solve the
  problem using numerical methods with the values provided below.
\end{asparaenum}

For both problems, plot the time dependence of heating, piston position,
pressure, and temperature assuming
\begin{subequations}\label{eq:num_values}
  \begin{align}
    L &= 20\un{m},    \\
    A &= 10^{-4}\un{m^2},    \\
M &= 2 \mul 10^{-3} \un{kg},
\\
f &= 9 \un{N},
\quad \text{corresponding to a pressure $f/A=9\mul 10^4 \un{Pa}$}, \\
    m &= 10^{-2} \un{kg},\\
    R &= 3 \mul 10^2\un{J/(kg\cdot K)}, \\
    c &= 7 \mul 10^2 \un{J/(kg\cdot K)},  \\ 
\zLa &= 5\mul 10^{-1}\un{m}, \\ 
    \zta &= 3 \un{s},
\\
\zacc&=-10 \un{m/s^2}\quad\text{(as in free fall)}.
  \end{align}
\end{subequations}
(If the value of $L$ appears impractically large, consider that this is an
idealization of the real set-up of fig.~\ref{fig:schufle}: the distance $L$
of the exercise is an effective distance given by $L = \zLh + \ztV/A$,
where $\zLh$ is the distance of the piston from the base of the real tube
and $\ztV$ is the volume of the vessel connected to the latter; the same
holds for $h$. The real tube can be between one and two metres long.)

\section{Solution and remarks}
\label{sec:solution}
Let us denote the gas' pressure, temperature, and internal energy by
$p(t)$, $\zT(t)$, $U(t)$. The equations relevant to the problem are:
\begin{asparaitem}
\item Momentum balance for the piston, upon which act the pressure of the
  gas $p(t)$ and the external force $f$:
  \begin{equation}
    \label{eq:momentum_i}
    Ap(t) -f -m\ddot{h}(t) =0,
  \end{equation}
where a dot stands for the time derivative: $\dot{h}\defd \di
h/\di t$, \etc
\item Energy balance for the gas:
  \begin{equation}
    \label{eq:energy_i}
    \dot{U}(t) = \zQ(t) - p(t) A \dot{h}(t),
  \end{equation}
  where $p(t) A \dot{h}(t)$ is the working done by the gas, since the
  volume is obviously given by $A h(t)$ and $A$ is constant. Note that
  $\zQ(t) = \dot{Q}(t)$, where $Q(t)$ is the integrated heating from some
  arbitrarily specified time until time $t$.
\item Constitutive law for the pressure of an ideal gas:
  \begin{equation}
    \label{eq:gaslaw_i}
    p(t)=\frac{\zM R \zT(t)}{A h(t)}.
  \end{equation}
And finally
\item constitutive equation for the gas' internal energy, as specified in
  the problem:
  \begin{equation}
    \label{eq:ienergy_i}
    U(t) = \zM c \zT(t).
  \end{equation}
\end{asparaitem}

The equations above constitute a system of coupled ordinary differential
equations for the unknown $h$, $\zT$, $p$, $U$, given $\zQ(t)$. The system
can be easily simplified. Substituting the expression for the
pressure~\eqref{eq:gaslaw_i} in the momentum balance~\eqref{eq:momentum_i}
we can express $\zT$ in terms of $h$:
\begin{subequations}\label{eq:sol}
  \begin{equation}
    \label{eq:sol_T_h}
    \zT(t) = \frac{1}{\zM R}\, h(t) [f+m\ddot{h}(t)],
  \end{equation}
  from which, substituting in~\eqref{eq:gaslaw_i} and~\eqref{eq:ienergy_i},
  we also obtain an expression for the pressure:
  \begin{equation}
    \label{eq:sol_p_h}
    p(t) = \frac{1}{A} [f+m\ddot{h}(t)],
  \end{equation}
  and the internal energy:
  \begin{equation}
    \label{eq:sol_U_h}
    U(t) = \frac{c}{R}\, h(t) [f+m\ddot{h}(t)].
  \end{equation}
\end{subequations}
From the latter two equations and the energy balance~\eqref{eq:energy_i} we
finally obtain an expression relating $h$ and $\zQ$:
\begin{equation}
  \label{eq:sol_Q_h}
    \zQ(t) =  \Bigl(1+\frac{c}{R}\Bigr)\dot{h}(t) [\zg+m\ddot{h}(t)] 
     +\frac{cm}{R} h(t)
     \dddot{h}(t).
\end{equation}
If the position function $h$ is given, the equation yields the heating
$\zQ$; if the heating is given, this is an ordinary differential equation
of third order from which $h$ can be obtained, possibly with numerical
methods, given appropriate initial conditions for $h$, $\dot{h}$, and
$\ddot{h}$.

Problem~\ref{item:findQ} is easily solved by substituting the
expression~\eqref{eq:h_question} for $h$ into~\eqref{eq:sol_Q_h}. We find
\begin{equation}
\zQ(t)=
\frac{6 c L\zLa m}{R \ztaa^3}+
\frac{3f \zLa (c+R)}{R  \zta^3} (t-\zti)^2+
\frac{6m \zLa^2 (4c+3R)}{R \zta^6} (t-\zti)^3.
\label{eq:Q_ass}
\end{equation}
Using equations~\eqref{eq:sol_T_h} and~\eqref{eq:sol_p_h} we also find the
time dependence of pressure and temperature:
  \begin{align}
    p(t) &= \frac{f}{A} + \frac{6 \zLa m}{A \ztaa^3} (t-\zti),
    \\
    \zT(t) &=
    \begin{multlined}[t][.8\columnwidth] 
\frac{f L} {M R}
+\frac{6 \zLa L m}{M R \zta^3} (t-\zti)
+\frac{f\zLa}{M R \zta^3} (t-\zti)^3 +{}\\
\frac{6 \zLa^2 m}{M R \zta^6} (t-\zti)^4.
    \end{multlined}
  \end{align}
These solutions are shown in fig.~\ref{fig:h1} for the numerical
values~\eqref{eq:num_values}. With those values, $\zQ$ is dominated by its
quadratic term in time, $\zT$ by its constant and cubic terms. 

\bigskip

Now to problem~\ref{item:findh}. Since the process is adiabatic,
$\zQ(t)=0$, the differential equation~\eqref{eq:sol_Q_h} can be rewritten as
\begin{equation}
  \label{eq:sol_0_h}
    \Bigl(1+\frac{c}{R}\Bigr)\dot{h}(t) [\zg+m\ddot{h}(t)] 
     +\frac{cm}{R} h(t)
     \dddot{h}(t) = 0,
\end{equation}
with the initial conditions
\begin{equation}
  \label{eq:initial_cond_0}
  h(\zti)=L,\qquad \dot{h}(\zti)=0,\qquad \ddot{h}(\zti)=\zacc.
\end{equation}
The differential equation above can be integrated once by rewriting it as
\begin{equation}
  \label{eq:sol_0_h_rewr}
\frac{m \dddot{h}(t)}{\zg+m\ddot{h}(t)}
=
-\Bigl(1+\frac{R}{c}\Bigr) \frac{\dot{h}(t)}{h(t)}
\end{equation}
and noting that each side is the derivative of a logarithm. Integrating
between $\zti$ and $t$, exponentiating the logarithms, rearranging, and
using the initial conditions~\eqref{eq:initial_cond_0} we obtain
\begin{equation}
  \label{eq:sol_0_h_integ}
[\zg+m\ddot{h}(t)]\, h(t)^\gamma
=
[\zg+m \zacc]\, L^\gamma
\end{equation}
with $\gamma \defd 1+R/c$, the ratio of specific heats of the gas.
Comparison with expression~\eqref{eq:sol_p_h} for the pressure shows that
this equation is the law for adiabatic processes, $p(t) [A h(t)]^\gamma =
p(\zti) [A h(\zti)]^\gamma$.

This differential equation does not seem to be solvable analytically, so we
proceed by numerical integration using the values~\eqref{eq:num_values}.
Note that these values could correspond to having the tube placed
vertically and suddenly dropping the piston in it at time $\zti$.

The numerical solution is shown in fig.~\ref{fig:h02}.
We observe an oscillatory behaviour of all quantities, with obviously the
same period of circa $0.8 \un{s}$. The oscillation amplitude of the piston
is around $1.5\un{dm}$, that of the pressure around $1\un{kPa}$, and that
of the temperature around $1 \un{K}$. One can show numerically that
changing the initial acceleration only changes the amplitudes of the
oscillations but not their period. 

Now let us see whether the differential equation~\eqref{eq:sol_Q_h} admits
any approximate form that can be solved analytically. If the inertial term
$m\ddot{h}(t)$ is small in comparison to the force $f$ and the
variations of $h(t)$ are small in comparison to its initial value $L$, \ie
\begin{equation}
  \label{eq:valid_appr}
  \frac{m\ddot{h}(t)}{f}\text{ and } \frac{h(t)-L}{L}\text{ small},
\end{equation}
then~\eqref{eq:sol_Q_h} can be approximated by
\begin{equation}
  \label{eq:sol_Q_h_appr}
    \zQ(t) =  \Bigl(1+\frac{c}{R}\Bigr)\zg\dot{h}(t)
     +\frac{cm L}{R} \dddot{h}(t).
\end{equation}
Integrating between $\zti$ and $t$ and defining
\begin{subequations}\label{eq:approx_harm}
  \begin{equation}
    \label{eq:h_redef}
    \zhh(t) \defd h(t) 
    - h(\zti) -\frac{cmL}{c+R}\ddot{h}(\zti),
\qquad
 Q(t) \defd \int_{\zti}^t \zQ(s)\,\di s
  \end{equation}
  we obtain
  \begin{equation}
    \label{eq:sol_Q_h_integr}
    Q(t) =  \Bigl(1+\frac{c}{R}\Bigr)\zg \zhh(t)
    +\frac{cm L}{R} \ddot{\zhh}(t).
  \end{equation}
\end{subequations}
This equation describes a driven harmonic motion and can be solved exactly
in terms of integrals of $Q$. The intrinsic period is
\begin{equation}
  \label{eq:period}
   \ztp=2\pu\sqrt{\frac{c m L}{(c+R) f}} = 0.78\un{s},
\end{equation}
which is approximately equal to that found numerically in the solution to
problem~\ref{item:findh}. In fact, that solution does reasonably satisfy
the conditions~\eqref{eq:valid_appr} and one can compare the numerical
result with those obtained using \eqn~\eqref{eq:sol_Q_h_integr}. For
$\zti\le t\le \zta$ it is found that the difference in the prediction of
$h(t)$ does not exceed $10^{-3}\; L$.

The solution found in problem~\ref{item:findh} can be experimentally
observed with an Assmann-\bd R\"uchardt apparatus
\citep{assmann1852,ruechardt1929} as modified by Schufle
\citey{schufle1957}, which is exactly our real set-up of
fig.~\ref{fig:schufle}. With this apparatus, the oscillatory behaviour we
found is used to calculate the ratio of specific heats of the gas, which in
our case is given by $\gamma=1+R/c$, by measuring the period $\ztp$
approximately given by \eqn~\eqref{eq:period}. The
values~\eqref{eq:num_values} are indeed approximately taken from Schufle's
article, and our numerical solution gives with good approximation the
period of oscillation observed in his experiments.
\figsol \fige \figdiff

It should be remarked however that for a smaller volume of gas, \ie\
smaller $L$, or bigger piston mass $m$, the
approximations~\eqref{eq:valid_appr} do not hold and the approximate
equation~\eqref{eq:sol_Q_h_appr} gives results that differ appreciably from
those of \eqn~\eqref{eq:sol_Q_h}. Assume \eg\ the values
\begin{subequations}\label{eq:num_values_new}
  \begin{align}
    L &= 1\un{m},    \\
    A &= 10^{-4}\un{m^2},    \\
M &= 2 \mul 10^{-3} \un{kg},
\\
f &= 19 \un{N},
\\
    m &= 1 \un{kg},\\
    R &= 3 \mul 10^2\un{J/(kg\cdot K)}, \\
    c &= 7 \mul 10^2 \un{J/(kg\cdot K)},  \\ 
\zacc&=-10 \un{m/s^2},
  \end{align}
\end{subequations}
which represent a vertical $1\un{m}$ tube with a piston of mass $1\un{kg}$
acted upon by atmospheric pressure. With these values
\eqns~\eqref{eq:sol_Q_h} and~\eqref{eq:sol_Q_h_appr} give distinctly
different solutions, as can be seen in fig.~\ref{fig:hd}: continuous blue
line for \eqn~\eqref{eq:sol_Q_h} and dashed orange line for
\eqns~\eqref{eq:sol_Q_h_appr}. Not only is the period of the piston's
motion different: the first solution is an anharmornic oscillation; the
second, harmonic. As expected, \eqn~\eqref{eq:sol_Q_h} can present more
interesting behaviours than just harmonic motion.

\section{Arguing for the time variable in classical thermodynamics}
\label{sec:timeTD}






\subsection{Advantages}
\label{sec:advant}

With the solution and analysis of our little textbook exercise I hope to
have provided one more piece of evidence that time has its appropriate
place in classical thermodynamics and thermodynamic courses. Its explicit
presence has many advantages:

First, it makes thermodynamics feel closer in spirit to mechanics and
electromagnetism, giving a more unified view of physics.

Second, it allow us to avoid any mention of differentials, especially
`inexact differentials', and therefore does not require notions from the
mathematics of differential forms. Moreover, the fundamental
laws~\eqref{eq:laws_time}, stated with respect to time, show why there
cannot be differentials related to heating and working: separately, they
are not the rates of \emph{change} of anything. (For a beautiful, rigorous,
and fruitful use of differential \emph{forms} in physics see Burke
\citey{burke1985_r1987,burke1995}, Bossavit \citey{bossavit1998b}, Hehl
\amp\ Obukhov \citey{hehletal2000}, van Dantzig
\citey{vandantzig1954,vandantzig1934}, Schouten \citey{schouten1951_r1989},
Truesdell \amp\ Toupin \citey[\chap~F]{truesdelletal1960}.)

Both points above allow, as a third advantage, an easier study of problems
in which thermal, mechanical, and electromagnetic phenomena appear
together. Think what would happen if the piston of our exercise had an
electric charge and the tube was in a magnetic field. This kind of problems
is often avoided because of the clash between the mathematics typically
used in thermodynamics and in mechanics, a clash that leads to ugly
mathematical gymnics. An example is given in Schufle's article
\citey{schufle1957}: he must derive the equation of a harmonic oscillator
in a set-up similar to that of our exercise. He uses differentials in
deriving the necessary relations between thermodynamic quantities, but then
he must combine these with mechanical ones, arriving at an expression with
the differential of a force. At that point he has to concede that the force
is related to the double time-derivative of position, revealing that all
preceding differentials were but time derivatives in disguise. Our solution
in \sect~\ref{sec:solution} was obtained in a mathematically seamless way
instead, owing to the presence of time in the basic thermodynamic
equations.

Fourth --- and this is a very important point not only for teachers and
students, but also for researchers --- it makes it possible to study more
interesting thermodynamic systems for which the pressure or other
quantities depend not only on volume and temperature but also \emph{on
  their rates of change}, say $p(V, \zT, \dot{V})$, as it happens \eg\ in
synovial fluids \citep{knightetal1983,hardyetal1995} or in pyroelectric
materials \citep{chynoweth1956}. The mathematics of differentials can
hardly handle such systems. On this generalization I recommend the very
lucid and simply written work of Samoh\'yl \citey{samohyl1987}, especially
\chap~II.

Fifth, it is full of physical and mathematical pedagogic potentials, as the
interesting third-order ordinary differential equation~\eqref{eq:sol_Q_h}
shows: it leads to a rich class of mathematical problems and physical
behaviours (the driven harmonic oscillator being only a special case),
providing also pedagogic material for laboratory classes.


\subsection{Common, weak objections}
\label{sec:objections}

In conversations  with colleagues on this topic two main kinds of objection
appear against a presentation of thermodynamics with an explicit presence
of time. I find both very weak:

The first objection is that thermodynamics would only be concerned about
equilibrium and relations between equilibria, and would have nothing to say
about rates of change. I should rather call such a theory
thermo\emph{statics}; \cf\ Eckart \citey{eckart1940a}. But why should we
limit ourselves to thermostatics, especially when experiment shows that the
application to time-dependent phenomena is justified as we have seen \eg\
in Schufle's experiment above? We do not limit ourselves to statics within
mechanics. Some people have even the extreme view that \emph{there is no
  proper science of the rate of change of thermal systems}: thermodynamics
can only be thermostatics. I do not give much weight to this view because
it is simply contrary to experience, experimental evidence, and practice.
According to this view all theories about non-equilibrium thermodynamics,
routinely taught and used by engineers and taught in some higher physics
courses, would be nonsense. That is preposterous.
(Some people even say temperature is defined only at equilibrium,
contradicting themselves each time they check a weather thermometer.)

The second objection is that an explicit time dependence and the
appearance of differential equations is unwarranted for high-school and
undergraduate students in their first years. But would not classical
mechanics present exactly the same problem? Yet, many mechanical problems,
like the harmonic oscillator or ballistic problems, can be and are
routinely approached and solved without the need to study the full theory
of differential equations. The same could surely be done with
thermodynamics. Moreover, is using non-intuitive and typically poorly
explained `inexact differentials', without explicitly introducing the
theory of differential forms, really easier than using rates of
change?

\subsection{Inadequacy of the standard terminology of classical thermodynamics}
\label{sec:inadequacy}

Part of the inertia against the time variable in thermodynamics comes
from the terminology taught in thermodynamic courses. Indeed, even those
who are keen on allowing the presence of time in classical thermodynamics
do so only after some verbal exorcisms. They say, \eg, that a process is
`quasi-static' or that the thermodynamic system `passes through a sequence
of equilibrium states', or other similar phraseologies. I find most of
these phraseologies misleading, counter-intuitive, and argue that we should
change them. Their inadequacy can be seen from different points of view:

First: In mechanics, electromagnetism, quantum theory we use particular
terminologies to indicate that specific equations and laws apply to a
problem. When we say that a spring is perfectly, linearly elastic, we mean
that Hooke's law $\bm{F} = -k\bm{x}$ can be applied. If a body is called
`rigid', it means that its motion can be fully described by the equations
$\bm{F} = m \ddot{\bm{x}}_\text{c}$ and $\bm{\mathsf{M}}_\text{c}=
\bm{\mathsf{J}} \dot{\bm{\omega}} + \bm{\omega} \times \bm{\mathsf{J}}
\bm{\omega}$, with constant $m$ and $\bm{\mathsf{J}}$. When we speak of an
ohmic resistor, we mean that the equation $\Delta V=r I$ applies with a
constant resistance $r$. Now, when in thermodynamics we say that a process
is `quasi-static' or that the system `passes through a sequence of
equilibrium states', we simply mean that an equation like $p(t)V(t)=M R
\zT(t)$ can be applied at all times. But while the terms `elastic',
`rigid', and even `ohmic' give the right ideas, `quasi-static' is
misleading and counter-intuitive: in our exercise we found visible rates of
change of $7.5 \un{dm/s}$, $5\un{kPa/s}$, and $5\un{K/s}$ (see
fig.~\ref{fig:h02}); do we really want to call them `quasi-static'?
Remember that `static' here means: with reference to the speed of sound,
$340\un{m/s}$ for air; hence a `quasi-static' process need not be static at
all in the common sense of the word. As for saying that a visibly changing
system is `passing through a sequence of equilibrium states', it sounds to
me like saying that a moving object is `passing through a sequence of
states of rest': it is at best an unnecessary and ugly phraseology, at
worst simply nonsense. Any mention of `equilibrium' is misleading when
thermodynamic quantities are clearly, visibly changing. Let us keep that
term, like `rest', for situations in which quantities are not changing in
time. Instead of mentioning `instantaneous equilibrium' or `quasi-static'
only to mean that the ideal-gas equation, \eg, can be used at all times,
let us say that the system is \emph{homogeneous at all times}: this implies
that it can be described at all times by a single density value, a single
pressure value, a single temperature value, \etc, instead of field
quantities as in continuum thermomechanics. 
Note in fact the analogy between a homogeneous body and a rigid body:
rigidity and homogeneity both imply that the field quantities describing
deformations in the full continuum-\bd thermo\-mechanical treatment can be
replaced by a smaller set of mechanical or thermal quantities valid for the
whole body.


Second: The argument typically adduced for the use of the term
`quasi-static' is that the system can be considered static --- in
equilibrium --- in comparison to \emph{some} time scale.
Yet, we do not call the Space Shuttle's motion at $7600\un{m/s}$
\citep{nasaarchive2011} `quasi-static' --- but perhaps we should, given
that the motion is in the province of Galileian-relativistic mechanics
where every body has to be almost at rest in comparison with the velocity
of light? Considering again the analogy between homogeneous and rigid
bodies: No body is perfectly rigid, and the equations for a rigid body make
sense only if the forces and torques do not exceed the internal ones that
keep the body together. Also, the action of a force on a part of the body
will not propagate instantly to the rest of the body: waves will propagate
in its interior, and they can be detected at appropriate space-time scales.
We usually do not point out all these facts when examining a rigid-body
problem. Analogously, no body is perfectly homogeneous, and its equations
make sense only when changes happen on a smaller scale than the velocity of
sound in the body; these changes propagate from point to point and could be
detected at appropriate space-time scales. We should feel the need to point
out these facts neither more nor less than we do in the case of the rigid
body.


Third: It is true that the construction and use of a thermometer or a
pressure gauge involves notions like response time and coupling (see \eg\
Quinn \citey{quinn1983_r1990}). But so does the construction of
position-gauges, velocity-gauges, and dynamometers. Some thermometers are
inappropriate when the temperature changes too rapidly, but the same is
true of velocity-gauges too. In mechanics and electromagnetics beginning
courses the theory of a physical phenomenon and the complex theory of
measurement behind it are usually kept quite separate. Positions,
accelerations, forces, charges, currents, fields are supposed to be
measurable with some appropriate instrument. The same should be true in
teaching thermodynamics; \cf\ Truesdell \citep[Lecture~1,
pp.~78--79]{truesdell1969_r1984}.

Fourth: In mechanics we enjoy the liberty to consider abstract situations
leaving unmentioned how they can be realized, in order to concentrate us
upon the equations and the phenomenology these exhibit. Springs exactly
obeying Hooke's law for arbitrary values of elastic constant and elongation
are supposed to exist for the sake of the mathematical problem. Thus it
ought to be in thermodynamics, but this liberty is rarely enjoyed there. In
mechanics we can say that a particle is in a harmonic potential, without
further question about how that is achieved (springs?\ some kind of
field?). Analogously, we should have the right to say, \eg, that the
temperature of a system is constant, without needing to mention `heath
baths' all the time. That the temperature is constant, or is changing in
some prescribed way, is all we need to know for our problem.

Truesdell makes these points very clear in his typical sharp and witty
style; I invite you to read the full book from which this quotation is taken:



  `For ten years or more, everyone studying materials has seen that both
  thermodynamics and mechanics must be brought to bear, but these sciences
  as hitherto taught do not mesh. Thus, if we take up almost any recent
  book with ``continuum mechanics'' or ``material science'' in its title,
  we find a chapter on thermodynamics, but that chapter presents a curious
  contrast with the same author's pages, earlier in the book, on pure
  mechanics. There the reader is faced by mappings, fields of vectors and
  tensors, Jacobians, differential invariants, perhaps even Christoffel
  symbols and affine connections; of course he is presumed familiar with
  calculus as calculus has been taught for the last fifty years. He can
  understand dynamical equations in tensorial form:
\begin{equation*}
  \dive \mathbf{T} + \rho\mathbf{b} = \rho\ddot{\mathbf{x}}
\quad\text{or}\quad
{T^{km}}_{{,}m} + \rho b^k = \rho\ddot{x}^k. 
\end{equation*}
He is informed, in each case, what the dependent and independent variables
are, he is presented with explicit differential equations and
boundary-value problems, he is shown many special solutions in concrete
cases and is directed to grand tomes where he can find thousands more such
solutions not given in the book he is reading, and often he is told about
some major problems still unsolved and is challenged to solve them himself.
The same reader of the same book then reaches the chapter on
thermodynamics, where he is faced with the ``axiom''
\begin{equation*}
  \zT\, dS \geqq \delta Q.
\end{equation*}
He is told that $dS$ is a differential, but not of what variables $S$ is a
function; that $\delta Q$ is a small quantity not generally a differential;
he is expected to believe not only that one differential can be bigger than
another, but even that a differential can be bigger than something which is
not a differential. He is loaded with an arsenal of words like piston,
boiler, condenser, heat bath, reservoir, ideal engine, perfect gas,
quasi-static, cyclic, nearly in equilibrium, isolated, universe---words
indeed familiar in everyday life, doubtless much more familiar than
``tangent plane'' and ``gradient'' and ``tensor'', which he learned to use
accurately and fluently in the earlier chapters, but words that never find
a place in the mathematical structure at all, words the poor student of
science is expected to learn to hurl for the rest of his life in a rhetoric
little sharper than a housewife's in the grocery store. The mathematical
structure, in turn, is slight. There are no general equations to be solved,
no boundary-value or initial-\bd value problems set, no general theorems
characterizing classes of solutions. The examples or exercises require no
more than calculating partial derivatives or integrals of given functions
or their inverses and plugging numbers into the results. The references
cited lead to other books containing just the same material, perhaps
otherwise explained and ordered, but no broader or clearer in concept, and
equally unmathematical. No problems, in the sense that the word ``problem''
has in the theories of mechanics or electromagnetism or optics or heat
conduction, are solved. Neither are any open problems stated. The reader
must presume that thermodynamics is an exhausted as well as exhausting
subject, with nothing left to be done'
\citep[Lecture~1, pp.~60--61]{truesdell1969_r1984}.


\begin{acknowledgements}
  Many thanks to Louise, Marianna, Miriam
  \langitalian{per i loro continui supporto e affetto}; to Ingemar
  Bengtsson and Giacomo Mauro D'Ariano for discussions; to Lucien Hardy and
  Lee Smolin for encouragement; to swing music; to R.\ Cheese; and to the
  developers and maintainers of \LaTeX, arXiv, Emacs, AUC\TeX, MiK\TeX,
  IrfanView, WinDjView, GIMP, Inkscape. Research at the Perimeter Institute
  is supported by the Government of Canada through Industry Canada and by
  the Province of Ontario through the Ministry of Research and Innovation.
\end{acknowledgements}



\defbibnote{prenote}{%
%
}

\newcommand{\citein}[2][]{\textnormal{\textcite[#1]{#2}}\addtocategory{1}{#2}}
\newcommand{\citebi}[1]{\textcite{#1}\addtocategory{1}{#1}}
\defbibfilter{1}{\category{1} \or \segment{1}}
\printbibliography[filter=1,prenote=prenote]


\end{document}